\documentclass[submitting]{nst}

\usepackage{subfigure,dcolumn}
\usepackage{epstopdf}
\usepackage{mhchem}
\usepackage{upgreek}

\begin{document}

\title{Charged hadron spectra and anisotropic flow from the AMPT model with shear viscous transport dynamics simulations at RHIC}

\author{Yao Zhang}
\email{yaozhang@jxust.edu.cn}
\affiliation{School of Energy and Mechanical Engineering, Jiangxi University of Science and Technology, Nanchang 330013, China}

\begin{abstract}
We utilized the AMPT model to simulate the shear viscous transport dynamics of parton matter in Au+Au collisions at a constant specific shear viscosity and varying phase transition temperatures at $\sqrt{s_{NN}}=200\rm{GeV}$. The resulting charged hadron spectra and anisotropic flow profiles correspond closely with experimental data. The transverse momentum spectra and longitudinal decorrelations are essentially unaffected by the phase transition temperature. An increase in the phase transition temperature leads to a rise in particle yields at midrapidity, accompanied by a decrease in both elliptic and triangular flows over a range of transverse momenta and pseudorapidities.
\end{abstract}


\maketitle

\section{Introduction}\label{sec.I}

The primary goals of contemporary high energy heavy-ion physics include pinpointing the critical point of phase transition within the QCD phase diagram and elucidating the transport characteristics of the quark-gluon plasma (QGP). At the BNL Relativistic Heavy-Ion Collider (RHIC) and the CERN Large Hadron Collider (LHC), heavy nuclei collisions with sufficient energy have resulted in the formation of a strongly coupled deconfinement QGP, which behaves as an almost perfect liquid with the smallest shear viscosity to entropy density ratio $\eta/s$ near the lower KSS limit \cite{1,2,3,4,5,6,7}. The specific shear viscosity has been incorporated into the viscous relativistic hydrodynamics simulations as an input parameter to obtain semi-quantitative constraints extracted from the measurements of particle spectra and anisotropic flow \cite{8,9,10,11}. The Bayesian parameter estimation method, applied to the viscous hydrodynamics and hadron transport hybrid model, has successfully provided precise estimates of the viscosity coefficient and its temperature dependence through comparative analysis with experimental data of various observables of QGP properties at LHC energy, indicating that the temperature dependence of shear viscosity increases with temperature and approaches the KSS bound near the phase transition temperature \cite{12,13,14}.

The AMPT model includes the transport process that simulates parton interactions and has a natural shear viscosity that is analytically or numerically calculable, dependent on the scattering cross section, in kinetic theory such as Israel-Stewart (IS) or Chapman-Enskog (CE) methods \cite{15}. For a fixed parton scattering cross section, the temperature dependence of the shear viscosity of parton matter estimated using the IS method in the AMPT model is opposite to that suggested by perturbative QCD and Bayesian analysis of the experimental data, which shows that the ratio of shear viscosity to entropy density increases as the temperature decreases \cite{16,17}. The relationship of shear viscosity with scattering cross section and temperature in IS method can be used to simulate the evolution process of partons with parameterized shear viscosity in the AMPT model \cite{18,19}. While the IS method is suited for isotropic scattering, partons in the AMPT model undergo forward-angle scattering, necessitating the use of the CE method that is better adapted for anisotropic scattering to estimate the shear viscosity \cite{15,20,21}. With the shear viscous transport dynamics simulation corrected by the CE method, we simulate Au+Au collisions at $\sqrt{s_{NN}}=200\rm{GeV}$ in the AMPT model with const $\eta/s$ which is less affected by temperature dependence at RHIC energy. The influences of different phase transition temperatures on the particle spectra and anisotropic flow of charged hadrons are studied. In Sec.\uppercase\expandafter{\romannumeral2}, we implement the shear viscous transport dynamics simulations of parton matter in the AMPT model. In Sec.\uppercase\expandafter{\romannumeral3}, the charged hadron spectra and anisotropic flow are discussed. We conclude in Sec.\uppercase\expandafter{\romannumeral4}.

\section{Shear viscous transport dynamics simulation in the AMPT model}\label{sec.II}

The AMPT model is a hybrid model for generating heavy ion collision events, which includes the initial conditions, parton interactions, hadronization, and hadronic interactions \cite{22}. Parton interactions are simulated by the ZPC parton cascade model based on non-equilibrium transport dynamics, which only includes two-body forward-angle scatterings \cite{23}.

In kinetic theory, the partonic scattering cross section determines the shear viscosity properties of the quark-gluon plasma and the Chapman-Enskog method for anisotropic scattering can express the shear viscosity as
\begin{equation}
\eta=\frac{4T}{5\sigma_{p} g(w)},
\end{equation}
where $\sigma_{p}$ is the partonic scattering cross section, $g(w)$ is the thermal average of $h(a)=4a(1+a)[(1+2a)\ln(1+\frac{1}{a})-2]$ and can be approximated as \cite{15,24}
\begin{equation}
g(w)\approx h(\frac{w^{2}}{v^{2}}),
\end{equation}
with $w=\mu/T$, $\mu$ is the screening mass and $v=11.31-4.847\exp(-0.1378w^{0.7338})$.

In the model, $\sigma_{p}$ can be approximated to $9\pi\alpha_{s}^{2}/(2\mu^{2})$ and the entropy density is $s=4g_{B}T^{3}/\pi^{2}$ by assuming the parton matter of massless three-flavor quarks at temperature $T$. The shear viscosity to entropy density ratio is \cite{21}
\begin{equation}
\eta/s\approx\frac{2\pi w^{2}}{45g_{B}\alpha_{s}^{2}g(w)},
\end{equation}
where $g_{B}=52$ and $\alpha_{s}$ is the QCD coupling constant.

According to Eq.(3), the specific shear viscosity $\eta/s$ depends on the ratio of $\mu/T$ for a fixed $\alpha_{s}$. Thus, by adjusting $\mu/T$, it is possible to simulate the partonic evolution process that complies with the parameterized $\eta/s(T)$ in the AMPT model. Under the Boltzmann distribution, the temperature of the parton phase can be estimated with the energy density, $T=(\frac{\pi^{2}\epsilon}{3g_{B}})^{\frac{1}{4}}$. Using Gaussian smearing approximation to smear each pointlike parton with a three-dimensional Gaussian distribution of its total energy allows for obtaining the distribution function of energy density at evolution time $t$, representable as \cite{25,26}
\begin{equation}
\epsilon(x,y,z)=\sum N_{i}\exp\Biggl[-\frac{(x-x_{i})^{2}+(y-y_{i})^{2}+(z-z_{i})^{2}}{2\sigma_{i}^{2}}\Biggl],
\end{equation}
where $N_{i}=(\frac{1}{2\pi})^{\frac{3}{2}}\frac{1}{\sigma_{i}^{3}}E_{i}$ provides the proper normalisation, $(x_{i},y_{i},z_{i})$, $\sigma_{i}$ and $E_{i}$ are the position vector, Gaussian width and energy of parton $i$. A fixed Gaussian width would lead to an overestimation of the energy density in the early stages of partonic evolution, therefore a Gaussian width, $\sigma_{i}\propto \frac{1}{2E_{i}}$, inversely related to the parton energy is adopted.

\begin{figure}
\includegraphics[scale=0.42]{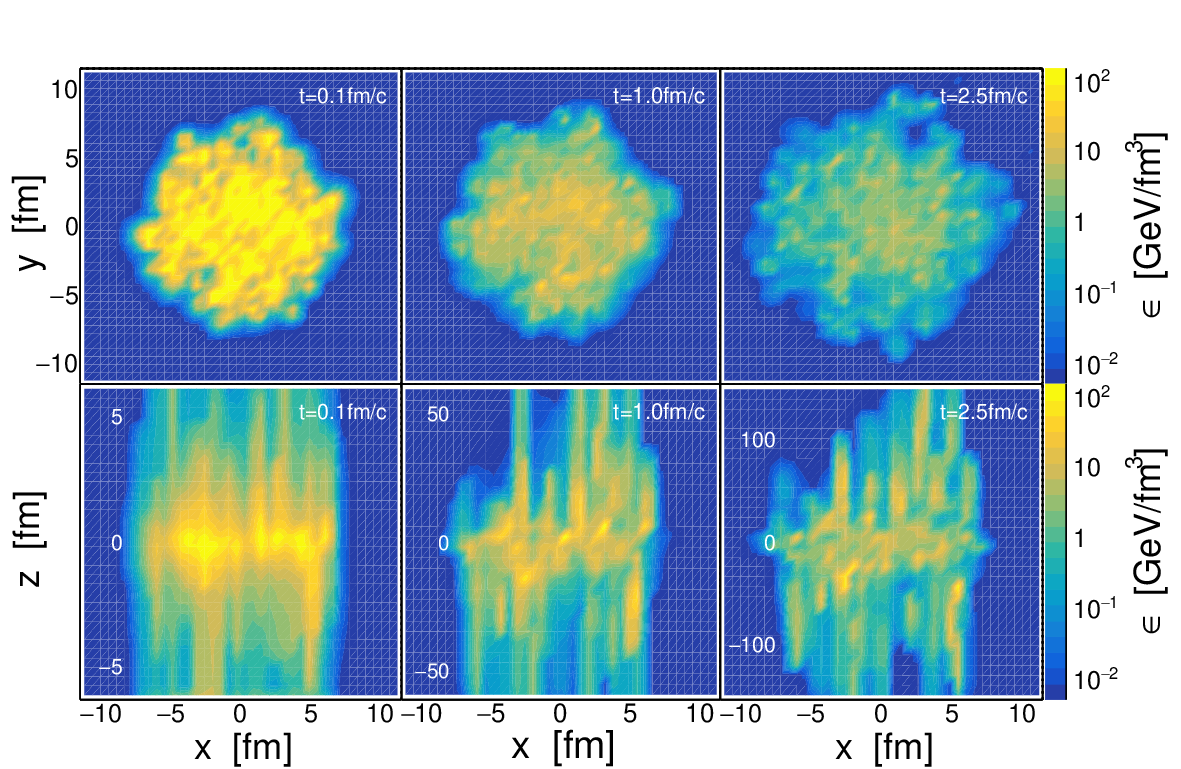}
\caption{\label{fig:1} Partonic energy density distribution in x-y plane (top) and x-z plane (bottom) for Au+Au central collisions at $\sqrt{s_{NN}}=200\rm{GeV}$ in the AMPT model: $t=0.1,1.0,2.5{\rm{fm/c}}$ from left to right.}
\end{figure}

Figure 1 shows the partonic energy density distribution of a single event in the $x$-$y$ and $x$-$z$ plane for Au+Au central collisions at $\sqrt{s_{NN}}=200\rm{GeV}$, where $z$ is the beam axis and $x$ is the in of plane axis. It can be seen from the figure that even with an insufficient number of partons in a single event, a continuous and consistent energy density distribution can still be obtained by the Gaussian smearing approximation, which contains the spatial dependence and fluctuations of parton matter. The spatiotemporal distribution of the parton phase temperature can be calculated from the energy density distribution with a time step of 0.1fm/c, where the temperature of each cell with the width of $\Delta x=0.2\rm{fm}$, $\Delta y=0.2\rm{fm}$ and space-time rapidity $\Delta\eta_{\rm{s}}=0.2$ is derived from the energy density at the volume center point of the cell.

\begin{figure}
\includegraphics[scale=0.3]{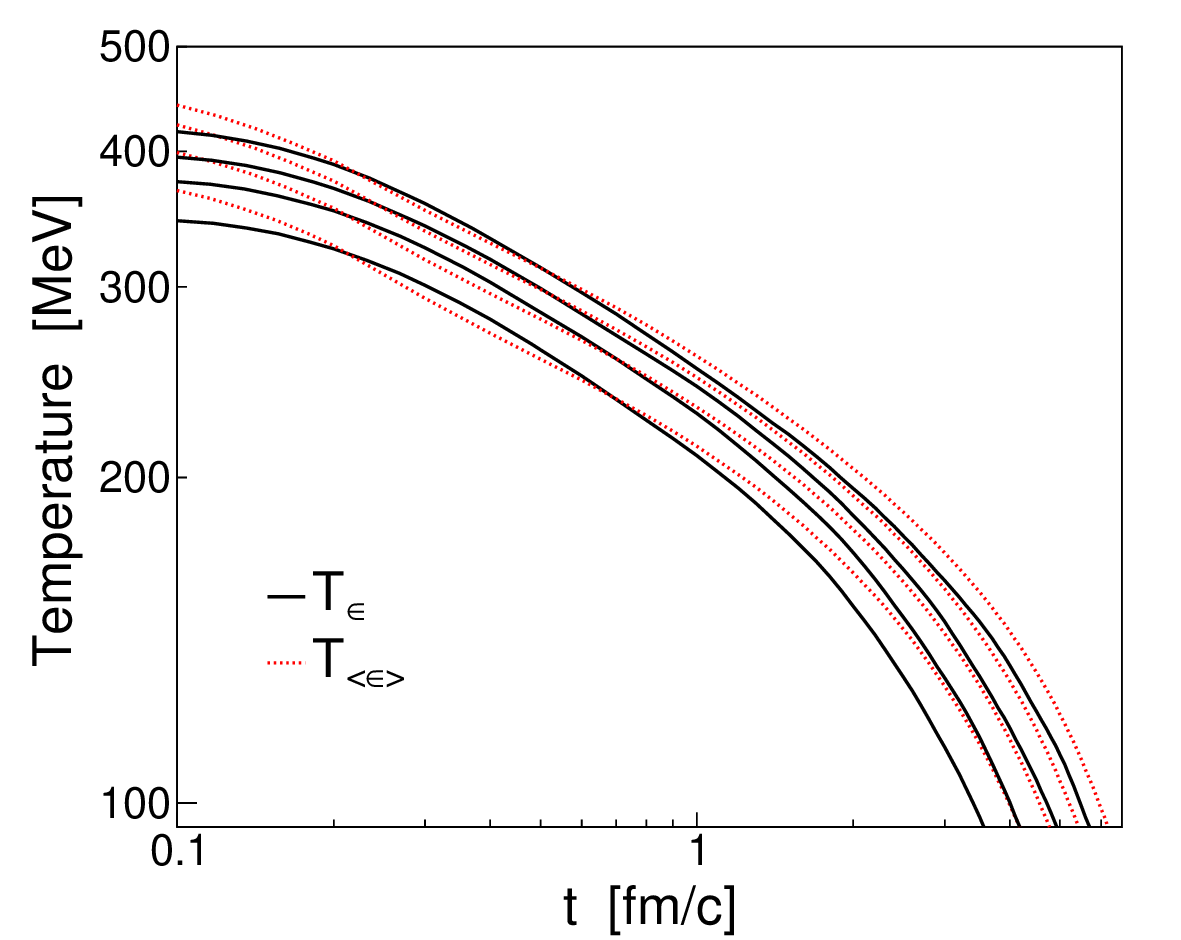}
\caption{\label{fig:2} The time evolutions of the effective temperatures $T_{\epsilon}$ of center point and $T_{<\epsilon>}$ of center cell at different centralities for Au+Au collisions at $\sqrt{s_{NN}}=200\rm{GeV}$. Curves from top to bottom correspond to 10-20$\%$, 20-30$\%$, 30-40$\%$, and 40-50$\%$ centralities, respectively.}
\end{figure}

Figure 2 shows the time evolutions of two effective temperatures averaging over events at different centralities for Au+Au collisions at $\sqrt{s_{NN}}=200\rm{GeV}$. The effective temperature $T_{\epsilon}$ and $T_{<\epsilon>}$ of parton matter are both calculated from the partonic energy density, with the difference being that the former uses Gaussian smearing approximation at the center point, while the latter is the average energy density of partons in the volume of center cell within transverse radius $r_{\rm{T}}<2\rm{fm}$ and space-time rapidity $|\eta_{\rm{s}}|<0.25$. The consistency of $T_{\epsilon}$ and $T_{<\epsilon>}$ over time indicates that the effective temperature $T_{\epsilon}$, obtained using Gaussian smearing approximation, can reflect the temperature evolution of the parton interactions. By utilizing the temperature distribution during the evolution of parton matter, quantitative control of parameterized specific shear viscosity in the model can be achieved by adjusting the screening mass which equates to altering the partonic scattering cross section in two-body scattering processes.

\section{Charged hadron spectra and anisotropic flow}\label{sec.III}

The shear viscous transport dynamics simulation can generates heavy ion collision events with specific parametrized shear viscosity $\eta/s$ of parton matter in the AMPT model. The temperature distribution obtained through Gaussian smearing approximation can also be used to examine the effects of phase transition temperature $T_{tr}$ on the measured observables within non-equilibration transport dynamic processes. When the effective temperature of the cell is below $T_{tr}$, the partons cannot undergo scattering in that cell and prepare for hadronization. The parameters of the Lund string fragmentation $a_{L}=0.55$ and $b_{L}=0.2\rm{GeV^{-2}}$, the coupling constant $\alpha_{s}=0.33$ are chosen for Au+Au collisions at $\sqrt{s_{NN}}=200\rm{GeV}$ in the model.

\begin{figure}
\includegraphics[scale=0.3]{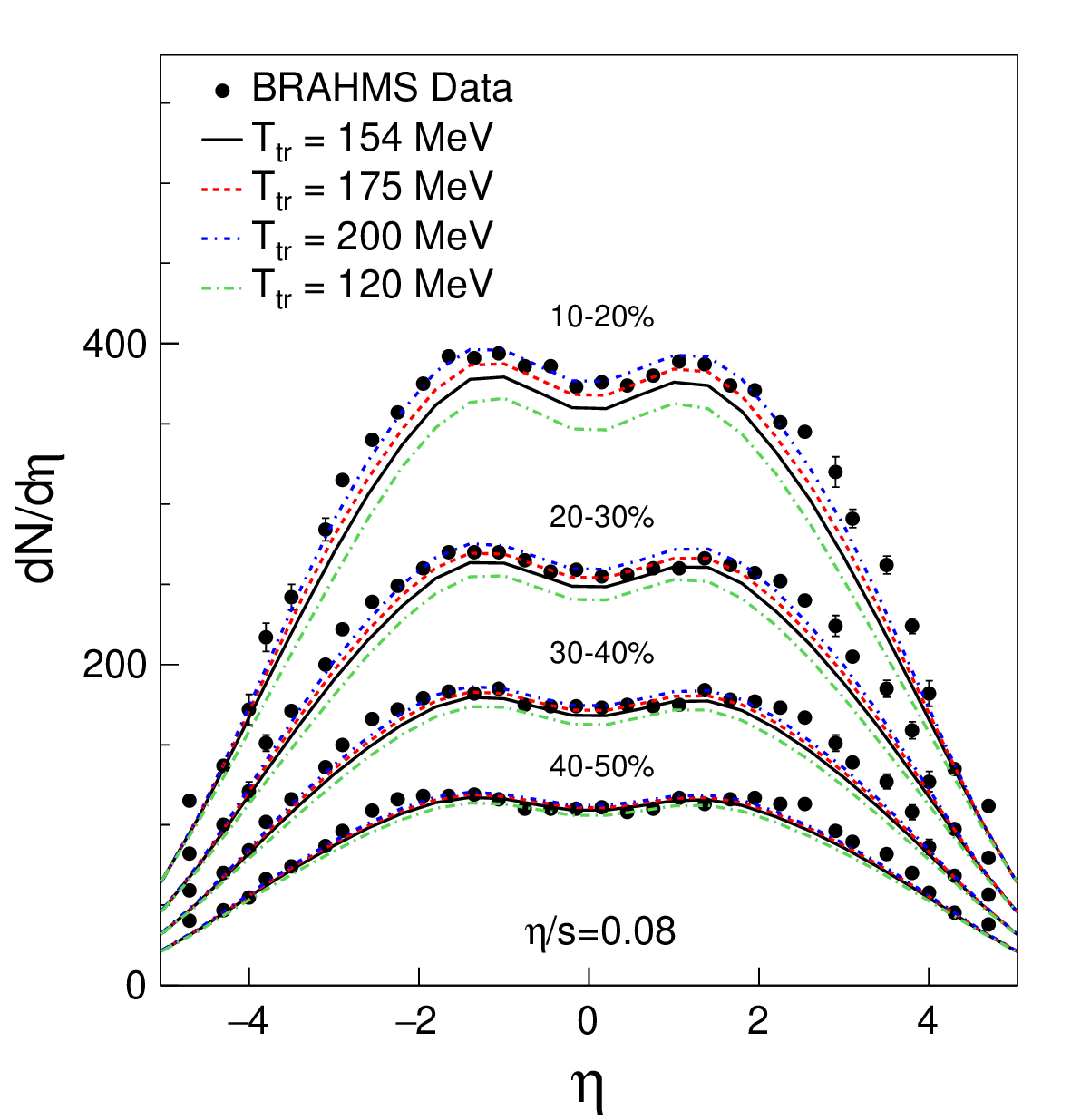}
\caption{\label{fig:3} The charged hadron pseudorapidity spectra for Au+Au collisions at $\sqrt{s_{NN}}=200\rm{GeV}$ with $\eta/s=0.08$ in different centrality classes compared to experimental data from the BRAHMS collaboration \cite{27}.}
\end{figure}

\begin{figure}
\includegraphics[scale=0.3]{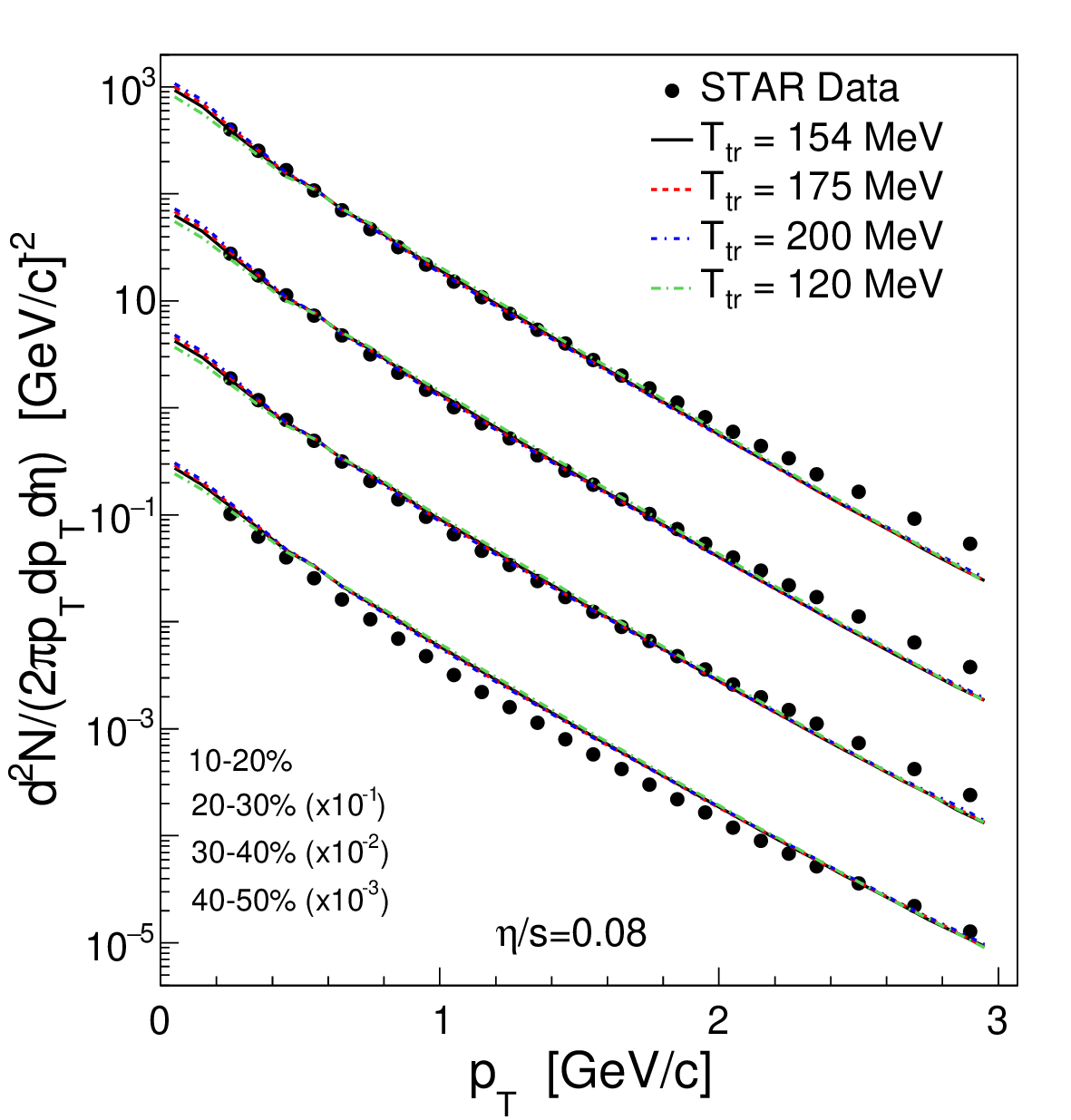}
\caption{\label{fig:4} Transverse momentum spectra of midpseudorapidity charged hadrons with $\eta/s=0.08$ in different centrality classes compared to experimental data from the STAR collaboration \cite{28}. Curves from top to bottom correspond to 10-20$\%$, 20-30$\%$, 30-40$\%$, and 40-50$\%$ centralities, respectively.}
\end{figure}

Figure 3 and figure 4 show the charged hadron pseudorapidity spectra and transverse momentum spectra, respectively, with the const $\eta/s=0.08$ for Au+Au collisions at $\sqrt{s_{NN}}=200\rm{GeV}$ compared to BRAHMS data and STAR data at various centralities. Both $dN/d\eta$ pseudorapidity distribution and $P_{T}$ spectra of charged hadron are well reproduced at all centralities near the KSS bound with the shear viscous simulation. Compared to the transverse momentum spectra, which are essentially unaffected by the phase transition temperature, a higher $T_{tr}$ tends to increase particle yields, especially in the midrapidity.

\begin{figure}
\includegraphics[scale=0.45]{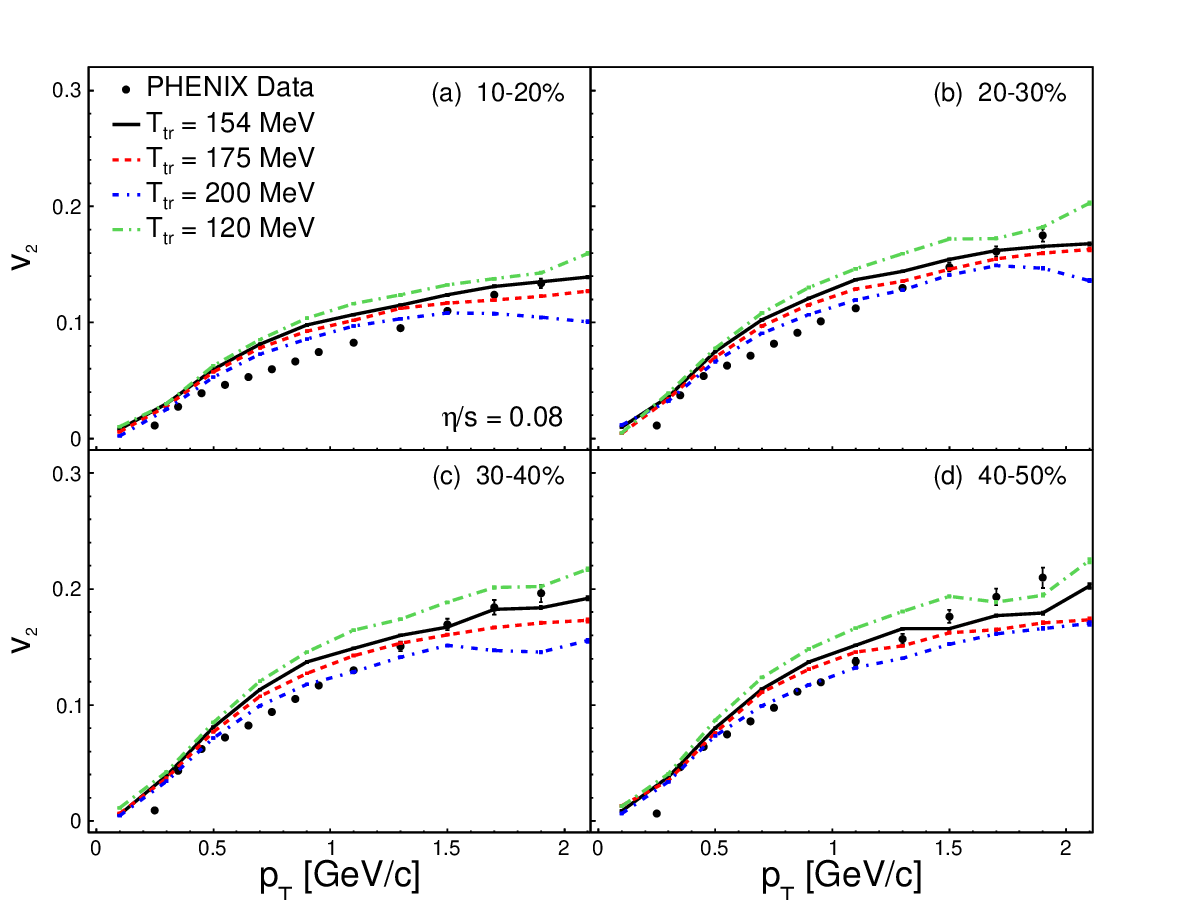}
\caption{\label{fig:5} Elliptic flow $v_{2}$ of midpseudorapidity charged hadrons varies with the transverse momentum with $\eta/s=0.08$ in different centrality classes compared to experimental data from the PHENIX collaboration \cite{31}.}
\end{figure}

\begin{figure}
\includegraphics[scale=0.45]{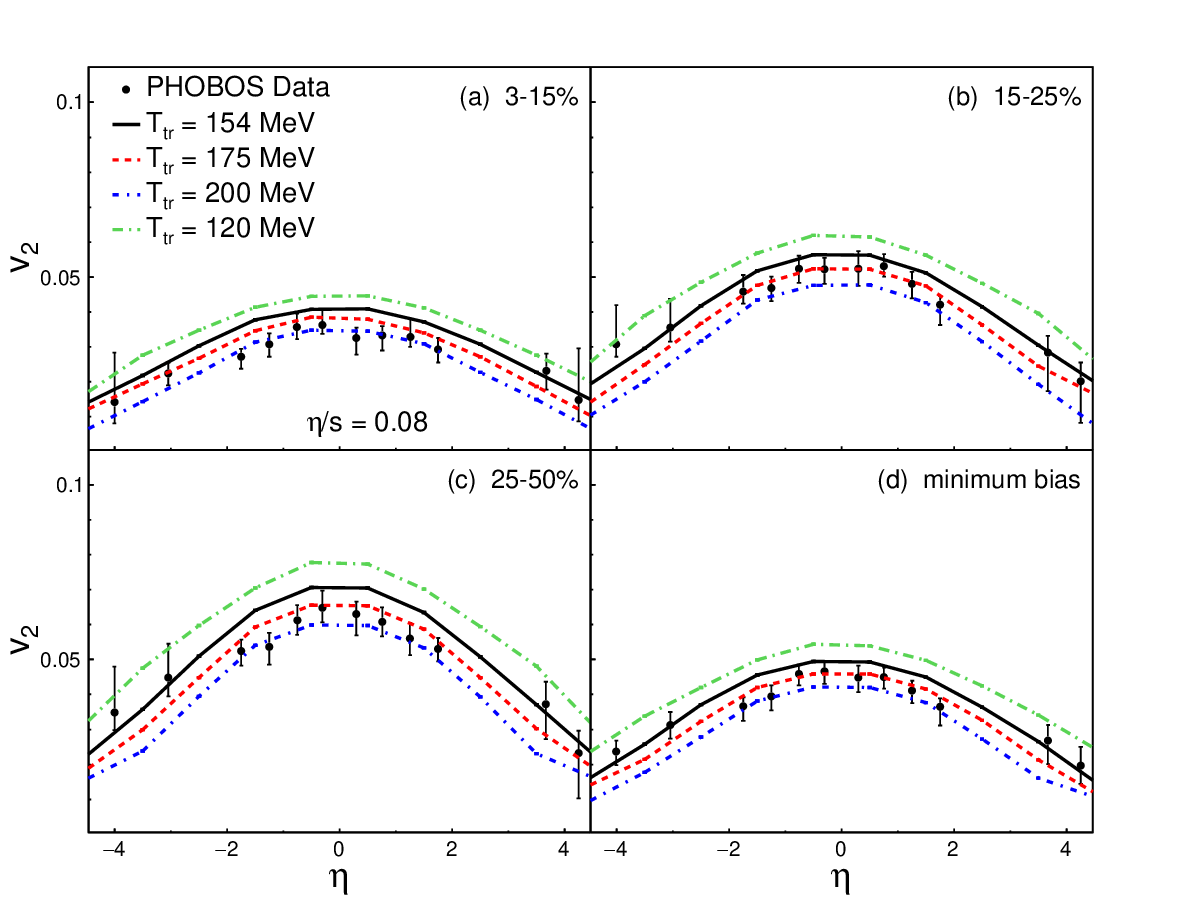}
\caption{\label{fig:6} Elliptic flow $v_{2}$ of charged hadrons varies with pseudorapidity with $\eta/s=0.08$ in different centrality classes compared to experimental data from the PHOBOS collaboration \cite{32}.}
\end{figure}

The anisotropic flow that develops from the pressure gradients due to the initial geometric shape of the collision is usually characterized by the flow harmonics of the particle azimuthal angle distribution, which can be calculated with the two-particle cumulant method as \cite{29,30}
\begin{equation}
v_{n}=\sqrt{\langle\mathrm{cos}n(\phi_{1}-\phi_{2})\rangle},
\end{equation}
where $\langle...\rangle$ denotes an average over all particle pairs in each event and then over all events, elliptic flow $v_{2}$ and triangular flow $v_{3}$ is the second and the third harmonic coefficient, which are sensitive to the initial geometry, transport properties and system evolution process.

Figure 5 and figure 6 show the charged hadron elliptic flow $v_{2}$ as a function of the transverse momentum and the pseudorapidity, respectively, compared to PHENIX data and PHOBOS data at various centralities. The elliptic flow can well describe the experimental data at different centrality classes. It is evident from the figures that the phase transition temperature inversely correlates with the value of elliptic flow, regardless of changes in transverse momentum or pseudorapidity. The higher $T_{tr}$, the smaller the corresponding $v_{2}$, and with changes in temperature, the elliptic flow value exhibits a consistent and uniform alteration along the longitudinal direction under the pseudorapidity distribution. With the same initial conditions and viscous properties, the system evolution directly affected by the phase transition temperature. The higher $T_{tr}$, the shorter the parton phase evolution time, and correspondingly, the longer the hadron phase evolution time. Compared to $\eta/s$ of parton matter near the KSS bound, the viscosity in the hadronic transport process is significantly increased. As $T_{tr}$ increases, the longer process of hadronic interactions enhances the suppression of $v_{2}$.

\begin{figure}
\includegraphics[scale=0.45]{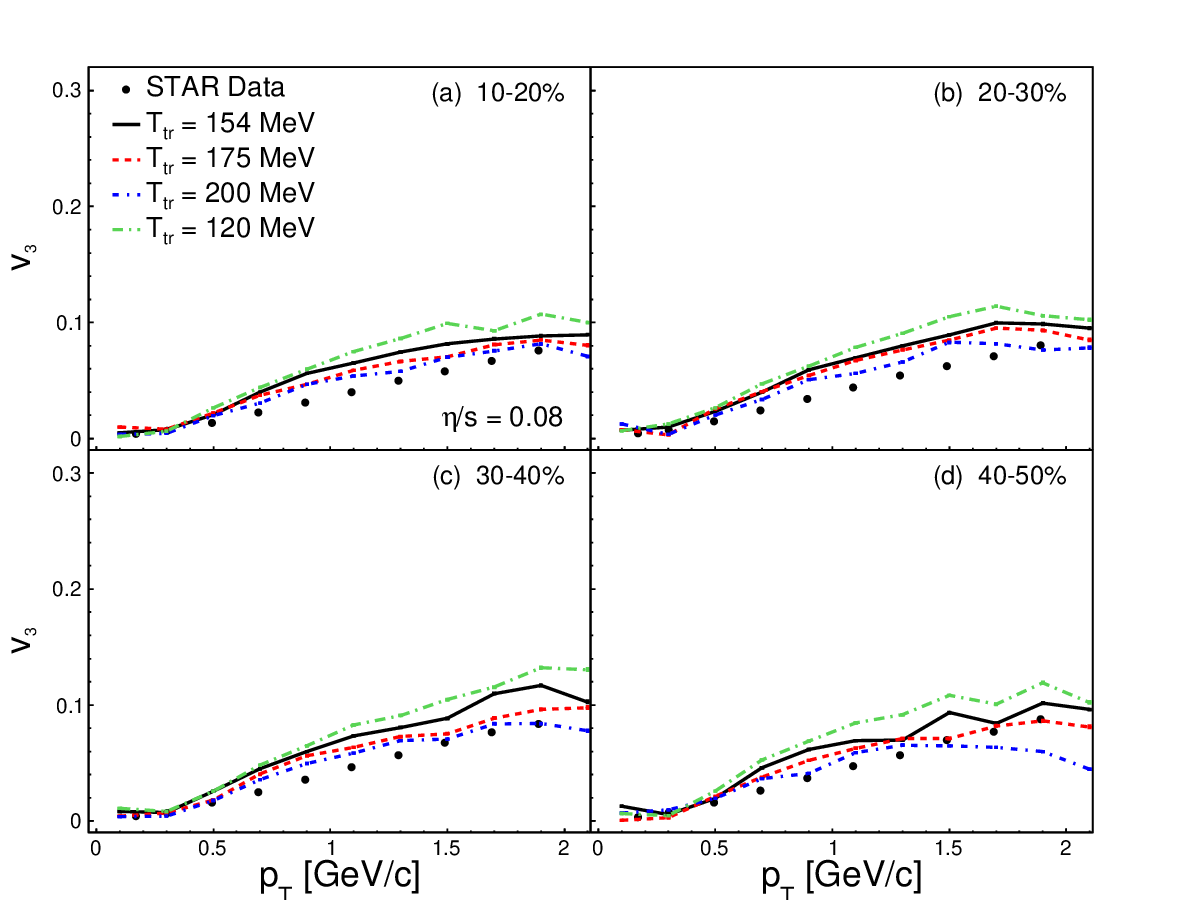}
\caption{\label{fig:7} Same as Fig.5 but for triangular flow $v_{3}$ compared to experimental data from the STAR collaboration \cite{33}.}
\end{figure}

\begin{figure}
\includegraphics[scale=0.45]{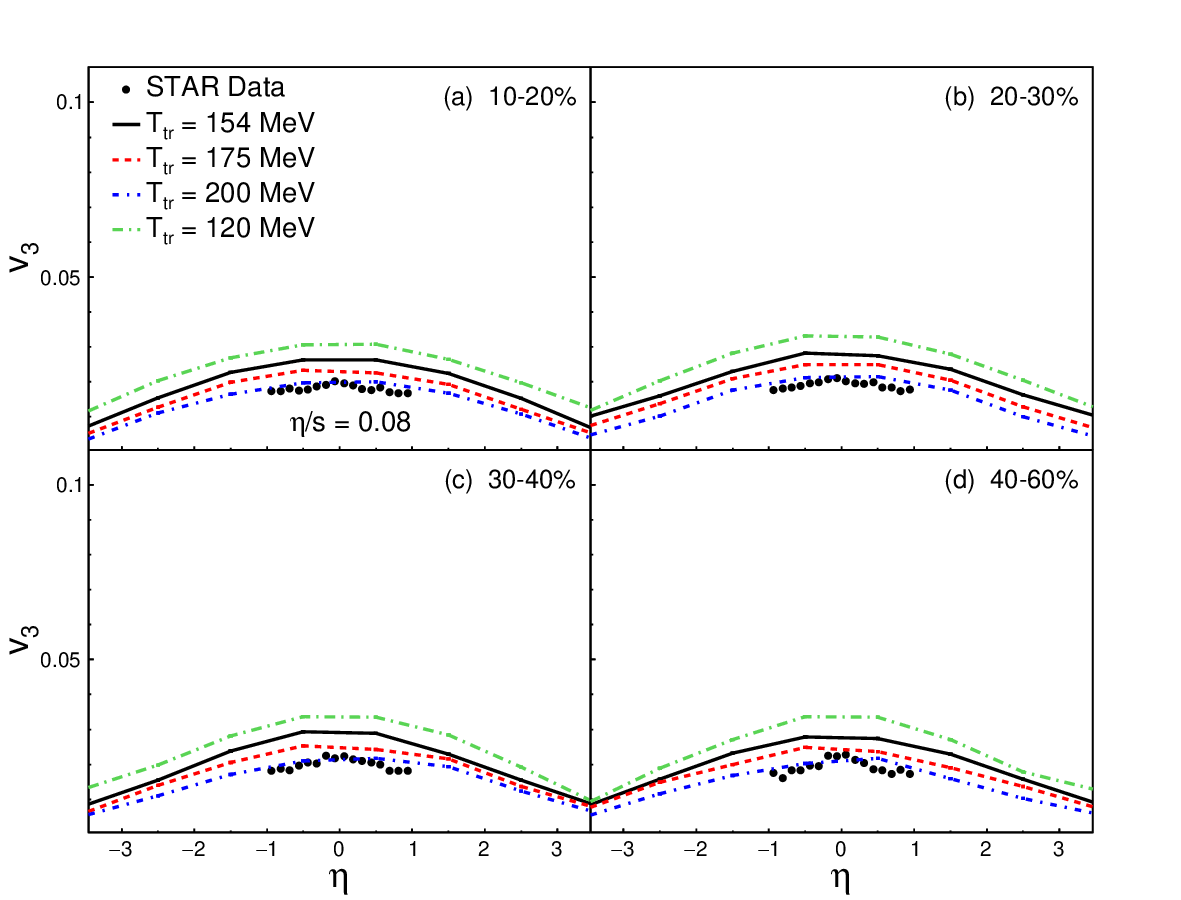}
\caption{\label{fig:8} Same as Fig.6 but for triangular flow $v_{3}$ compared to experimental data from the STAR collaboration \cite{33}.}
\end{figure}

Figure 7 and figure 8 show the charged hadron triangular flow $v_{3}$ as a function of the transverse momentum and the pseudorapidity compared to STAR data. The triangular flow provides a reasonable description of the experimental data and is also less sensitive to the centrality. $v_{3}$ calculated from the two-particle azimuthal correlation is different from the event plane method used in the experimental results, which is the main reason for the overestimation of flow. The triangular flow, which originates from the event-by-event fluctuations of the initial collision geometry, decreases with the increase of the phase transition temperature, just like the elliptical flow. The consistency in the effect of the phase transition temperature on $v_{2}$ and $v_{3}$ reflects the obvious influence of partonic transport dynamic evolution on the transfer from the initial geometric shape to momentum space, and this influence is manifested in the final anisotropic flow.

\begin{figure}
\includegraphics[scale=0.45]{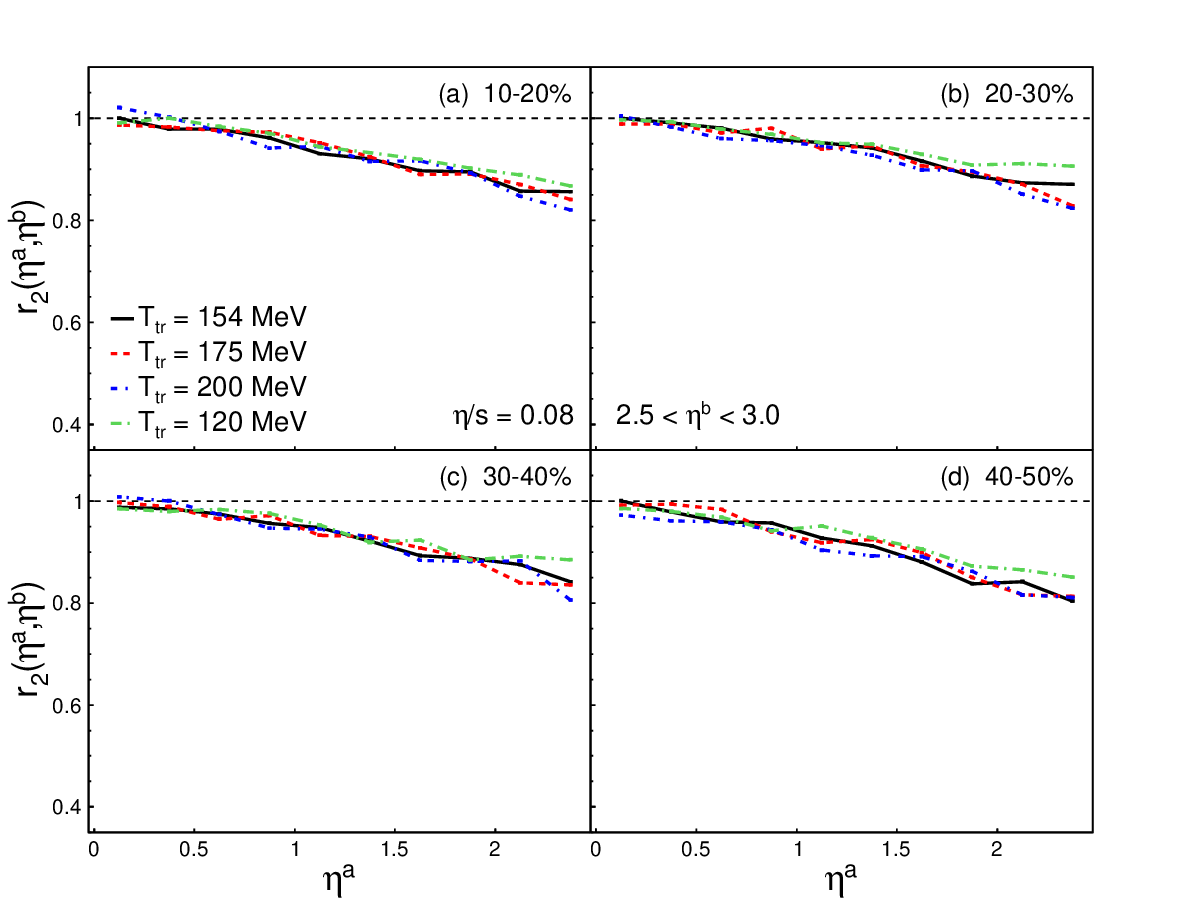}
\caption{\label{fig:9} The factorization ratio $r_{2}$ as a function of $\eta^{a}$ for $2.5<\eta^{b}<3.0$ with $\eta/s=0.08$ in different centrality classes.}
\end{figure}

\begin{figure}
\includegraphics[scale=0.45]{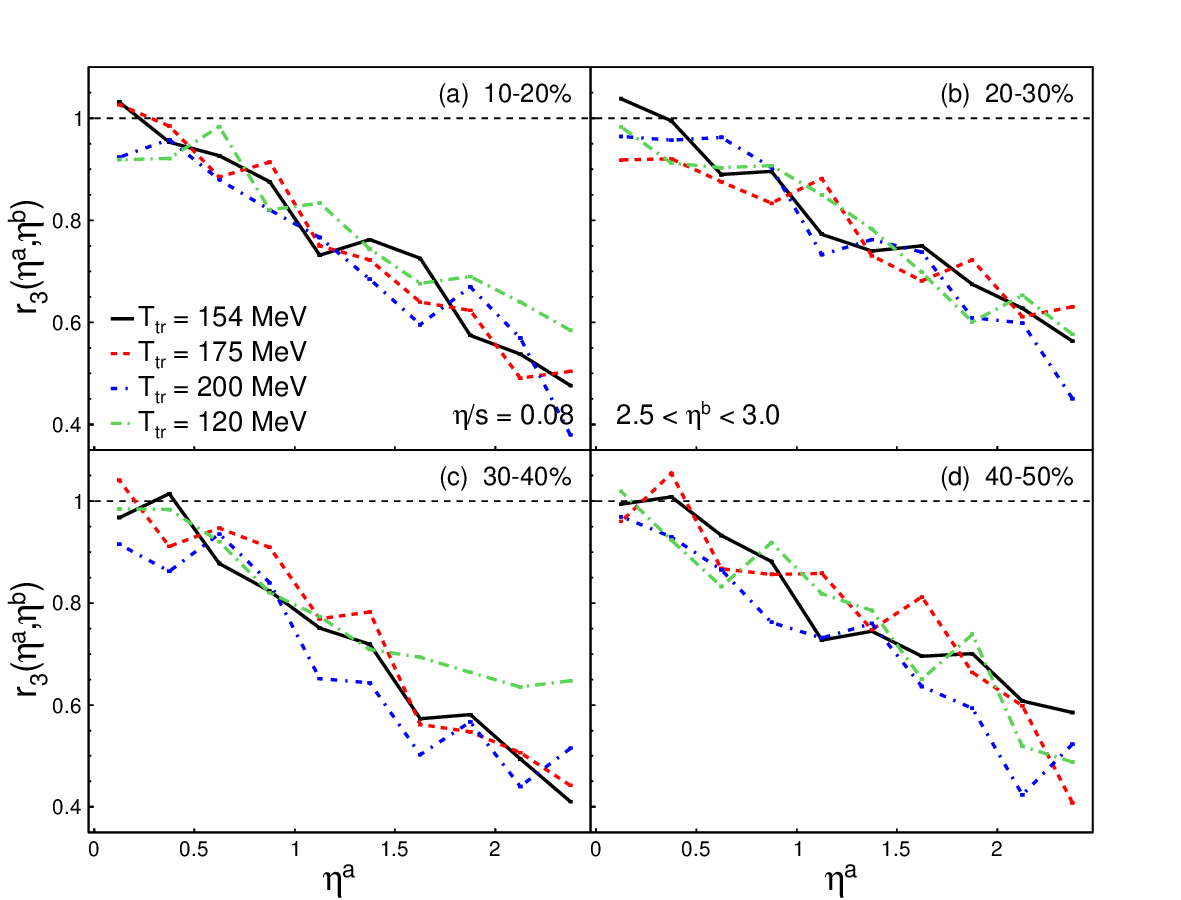}
\caption{\label{fig:10} Same as Fig.9 but for $r_{3}$.}
\end{figure}

The AMPT model exhibits a natural longitudinal correlation, and the transverse spatial geometry of the initial partonic matter is strongly correlated with the event plane within a finite pseudorapidity range, forming the longitudinal correlation of anisotropic flow through partonic and hadronic interactions. The longitudinal decorrelation of anisotropic flow can be used to explore the non-boost-invariant nature of the initial-collision geometry and final-state collective dynamics. The pseudorapidity-dependent factorization ratio, which represents the magnitude of decorrelation, is sensitive to the event-by-event fluctuations of the initial condition in the longitudinal direction and can be defined as \cite{33}
\begin{equation}
r_{n}(\eta^{a},\eta^{b})=\frac{\langle v_{n}(-\eta^{a})v_{n}(\eta^{b})\mathrm{cos}n(\Psi_{n}(-\eta^{a})-\Psi_{n}(\eta^{b}))\rangle}{\langle v_{n}(\eta^{a})v_{n}(\eta^{b})\mathrm{cos}n(\Psi_{n}(\eta^{a})-\Psi_{n}(\eta^{b}))\rangle}.
\end{equation}
Figure 9 and figure 10 show the factorization ratio $r_{2}$ and $r_{3}$ as a function of $\eta^{a}$ for $2.5<\eta^{b}<3.0$ at various centralities. Compared to the anisotropic flow which is affected by the phase transition temperature, the factorization ratios for $n=2$ and $3$ essentially remain unchanged with the variation of $T_{tr}$ at different centrality classes, indicating that $T_{tr}$ is unrelated to the non-boost-invariant nature of the collision's space-time evolution in the longitudinal direction.

\section{Summary} \label{sec.IV}

Based on the AMPT model with the shear viscous transport dynamics simulation of parton matter, we have studied Au+Au collisions with a const $\eta/s=0.08$ and various $T_{tr}$ at $\sqrt{s_{NN}}=200\rm{GeV}$, which can well describe the experimental results of charged hadron spectra and anisotropic flow. Relative to the transverse momentum spectra and longitudinal decorrelations that are essentially unaffected by the phase transition temperature, a higher $T_{tr}$ tends to increase particle yields at midrapidity and reduce the anisotropic flow value of $v_{2}$ and $v_{3}$. Elevating the phase transition temperature consistently suppresses the anisotropic flow across variations in transverse momentum and pseudorapidity, indicating a direct impact on the system evolution by shortening the parton phase and extending the hadron phase. The longitudinal decorrelation of anisotropic flow, represented by the factorization ratio, is insensitive to $T_{tr}$, suggesting that the phase transition temperature does not affect the non-boost-invariant nature of the collision's space-time evolution. The results aid in understanding the evolution of heavy ion collisions within RHIC and LHC, allowing for further investigation into the effects of phase transition temperature on measured observables using data across various energy regions.


\begin{thebibliography} {99}

\bibitem{1} I. Arsene et al. (BRAHMS Collaboration), Nucl. Phys. A {\bf757}, 1 (2005)
\bibitem{2} B. B. Back et al. (PHOBOS Collaboration), Nucl. Phys. A {\bf757}, 28 (2005)
\bibitem{3} K. Adcox et al. (PHENIX Collaboration), Nucl. Phys. A {\bf757}, 184 (2005)
\bibitem{4} J. Adams et al. (STAR Collaboration), Nucl. Phys. A {\bf757}, 102 (2005)
\bibitem{5} P. K. Kovtun, D.T. Son, A.O. Starinets, Phys. Rev. Lett. {\bf94}, 111601 (2005)
\bibitem{6} U. Heinz, R. Snellings, Annu. Rev. Nucl. Part. Sci. {\bf63}, 123 (2013)
\bibitem{7} E. Shuryak, Rev. Mod. Phys. {\bf89}, 035001 (2017)
\bibitem{8} C. Shen, U. Heinz, P. Huovinen, H. Song, Phys. Rev. C {\bf82}, 054904 (2010)
\bibitem{9} H. Song, S. A. Bass, U. Heinz, T. Hirano, C. Shen, Phys. Rev. Lett. {\bf106}, 192301 (2011)
\bibitem{10} E. Molnar, H. Holopainen, P. Huovinen, H. Niemi, Phys. Rev. C {\bf90} 044904 (2014)
\bibitem{11} R. A. Lacey, A. Taranenko, J. Jia et al, Phys. Rev. Lett. {\bf112}, 082302 (2014)
\bibitem{12} J. E. Bernhard, J. S. Moreland, S. A. Bass, Nat. Phys. {\bf15}, 1113 (2019)
\bibitem{13} D. Everett et al. (JETSCAPE Collaboration), Phys. Rev. Lett. {\bf126}, 242301 (2021)
\bibitem{14} J. E. Parkkila, A. Onnerstad, D. J. Kim, Phys. Rev. C {\bf104}, 054904 (2021)
\bibitem{15} N. M. MacKay, Z. W. Lin, Eur. Phys. J. C {\bf82}, 918 (2022)
\bibitem{16} J. Xu, C. M. Ko, Phys. Rev. C {\bf83}, 034904 (2011)
\bibitem{17} D. X. Wei, X. G. Huang, L. Yan, Phys. Rev. C {\bf98}, 044908 (2018)
\bibitem{18} Y. Zhang, J. Zhang et al., Phys. Rev. C {\bf96}, 044914 (2017)
\bibitem{19} Y. Zhang, J. Zhang et al., J. Phys. G: Nucl. Part. Phys. {\bf46}, 055101 (2019)
\bibitem{20} Z. W. Lin, L. Zheng, Nuclear Science and Techniques, {\bf32}, 113 (2021)
\bibitem{21} Y. Zhang, Q. Liu, Eur. Phys. J. A {\bf59}, 296 (2023)
\bibitem{22} Z. W. Lin, C. M. Ko, B. A. Li et al, Phys. Rev. C {\bf72}, 064901 (2005)
\bibitem{23} B. Zhang, Comput. Phys. Commun. {\bf109}, 193 (1998)
\bibitem{24} A. Wiranata, M. Prakash, Phys. Rev. C {\bf85}, 054908 (2012)
\bibitem{25} J. Steinheimer, M. Bleicher, H. Petersen et al, Phys. Rev. C {\bf77}, 034901 (2008)
\bibitem{26} L. G. Pang, Q. Wang, X. N. Wang, Phys. Rev. C {\bf86}, 024911 (2012)
\bibitem{27} I. G. Bearden et al. (BRAHMS Collaboration), Phys. Rev. Lett. {\bf88}, 202301 (2002)
\bibitem{28} J. Adams et al. (STAR Collaboration), Phys. Rev. Lett. {\bf91}, 172302 (2003)
\bibitem{29} N. Borghini, P. M. Dinh, J. Y. Ollitrault, Phys. Rev. C {\bf64}, 054901 (2001)
\bibitem{30} A. Bilandzic, R. Snellings, S. Voloshin, Phys. Rev. C {\bf83}, 044913 (2011)
\bibitem{31} S. Afanasiev et al. (PHENIX Collaboration), Phys. Rev. C {\bf80}, 024909 (2009)
\bibitem{32} B. Alver et al. (PHOBOS Collaboration), Phys. Rev. C {\bf83}, 024913 (2011)
\bibitem{33} L. Adamczyk et al. (STAR Collaboration), Phys. Rev. C {\bf88}, 014904 (2013)
\bibitem{34} V. Khachatryan et al. (CMS Collaboration), Phys. Rev. C {\bf92}, 034911 (2015)

\end{thebibliography}
\end{document}